\documentclass[9pt]{article}
\usepackage{INTERSPEECH2019}
\usepackage{hyperref}
\usepackage[super]{nth}
\usepackage{xcolor,colortbl}

\title{Can Speaker Augmentation Improve Multi-Speaker End-to-End TTS?}
\name{  Erica Cooper$^{1}$$^{\star}$\thanks{$\star$ Equal contribution. 
}, Cheng-I Lai$^{2}$$^{\star}$, Yusuke Yasuda$^{1}$, Junichi Yamagishi$^{1}$}

			\address{$^{1}$National Institute of Informatics, Japan $^{2}$Massachusetts Institute of Technology, USA }
			\email{$\{$ecooper,yasuda,jyamagis$\}$@nii.ac.jp, clai24@mit.edu}

\begin{document}

\maketitle
\begin{abstract}
    Previous work on speaker adaptation for end-to-end speech synthesis still falls short in speaker similarity. We investigate an orthogonal approach to the current speaker adaptation paradigms, \textit{speaker augmentation}, by creating artificial speakers and by taking advantage of low-quality data. The base Tacotron2 model is modified to account for the channel and dialect factors inherent in these corpora. In addition, we describe a warm-start training strategy that we adopted for Tacotron2 training. A large-scale listening test is conducted, and a distance metric is adopted to evaluate synthesis of dialects. This is followed by an analysis on synthesis quality, speaker and dialect similarity, and a remark on the effectiveness of our speaker augmentation approach.  Audio samples are available online\footnote{\href{https://nii-yamagishilab.github.io/samples-multi-speaker-tacotron/augment.html}{https://nii-yamagishilab.github.io/samples-multi-speaker-tacotron/augment.html}}.
    
\end{abstract}
\noindent\textbf{Index Terms}: Speaker augmentation, Speech synthesis, dialect identification, channel modeling, transfer learning

\section{Introduction}

Recent advances in end-to-end text-to-speech (TTS) synthesis enable the production of synthetic speech of high quality and good speaker similarity \cite{wang2017tacotron,ping2018clarinet,shen2018natural,park2019multi}. Although the speech quality approaches human naturalness, challenges still remain: first, to model many speakers simultaneously using a common model (termed ``multi-speaker TTS'') and second, to adapt to voices of arbitrary new speakers while minimizing the amount of data to be collected and requiring little or no additional model training (termed ``speaker adaptation'').

Previous work on speaker adaptation can be categorized into one of two general approaches. The first approach is simple fine-tuning  \cite{arik2018neural,kons2019high,chen2019sample}: the TTS model receives a small amount of additional training with target speaker data, which must be transcribed. The second approach is the use of external speaker embeddings \cite{jia2018transfer,cooper2020zeroshot}, which are extracted from separately trained automatic speaker verification (ASV) models, and the embeddings are input as speaker information to TTS models. 
This approach does not require transcriptions, and the speaker embedding can be computed from only a few utterances. However, it is reported that speaker similarity of unseen speakers is relatively low \cite{jia2018transfer}. On the other hand, there are a few attempts to use low-quality recordings for TTS. In \cite{hu2019neural}, low-quality recordings were used for fine-tuning based speaker adaptation.  Variational autoencoder based clean speech and noise factorisation \cite{8683561,hsu2019hierarchical} was also proposed for Tacotron TTS. They conducted speaker adaptation using artificially corrupted speech data \cite{8683561} or real noisy speech \cite{hsu2019hierarchical} and tried to create speaker-adapted `clean' TTS voices via the proposed factorisation.

In our previous work \cite{cooper2020zeroshot}, we constructed a multi-speaker Tacotron TTS model on the VCTK corpus \cite{vctk}, using speaker embeddings that are transferred from a separately trained ASV model, and performed zero-shot speaker adaptation.  The VCTK dataset contains high-quality speech recordings from around a hundred speakers of different English dialects. However, our model was overfitted to seen speakers, and voice characteristics and dialects of unseen speakers were not well reproduced, although the quality of synthetic speech was high \cite{cooper2020zeroshot}. We hypothesized that this number of speakers may be small for our task, and that increasing the number of training speakers can provide better coverage of the speaker space, avoid overfitting to seen speakers, and thus improve similarity and perceived dialects of unseen speakers.  However, TTS-quality datasets larger than VCTK are not easily found or created.  
 
A more realistic solution would be \textbf{\textit{speaker augmentation}}, that is, data augmentation for increasing the number of speakers used for neural network training. This has been investigated for ASV \cite{yamamoto2019speaker}, wherein they created ``artificial" speakers by simply re-sampling the original audio. They found that this approach improved their speaker models, and also that their system identified the artificial speakers as separate from the original ones.  This is known as ``vocal tract length perturbation'' (VTLP) and it also improved ASR \cite{jaitly2013vocal}. This could be useful for multi-speaker TTS since by adding more speakers, we can hope that neural networks will be aware of more diverse speaker characteristics and thus avoid overfitting to seen speakers. 

In addition to the above artificial speaker augmentation, we also consider another idea for speaker augmentation wherein we use non-ideal TTS data, that is, audio recordings that were collected for purposes other than TTS, and may not meet our usual high-quality recording standards, but have a larger number of speakers.  
%
%
However, carelessly mixing in data from worse recording conditions is expected to degrade the quality of synthesized speech. Furthermore, unlike artificial speaker augmentation, it also increases the number of different dialects included in the training database. We therefore once again borrow ideas from speaker recognition like the neural speaker embeddings in our previous paper, and propose an improved Tacotron speech synthesizer to explicitly handle the two factors, \textbf{channel} and \textbf{dialect}. Here, the channel is a factor jointly caused by frequency characteristics of recording equipment, noise and reverberation.  More precisely, in the proposed synthesizer, neural dialect embedding vectors are used to condition Tacotron's encoder, and channel labels are used to condition Tacotron's postnet. 


\section{Speaker Augmentation for TTS}

\begin{figure}[tb] 
\centering
\includegraphics[width=1.0\linewidth]{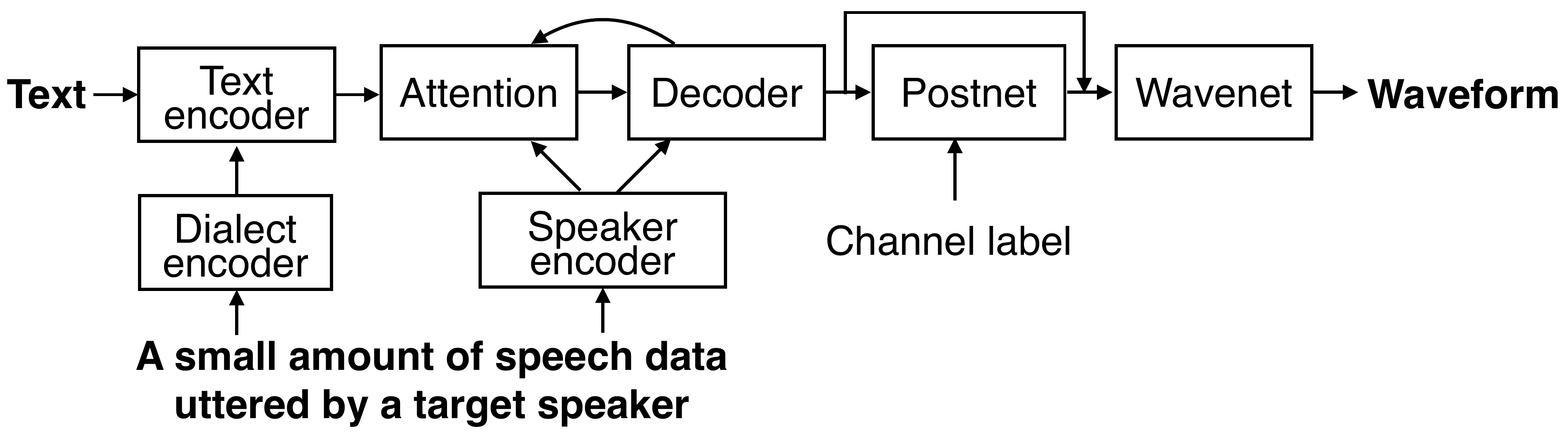}
\vspace{-3mm}
\caption{Diagram of our TTS system for speaker augmentation using low-quality data. This is modified from~\cite{cooper2020zeroshot} to accommodate for the additional channel and dialect factors, by adding a channel-aware postnet and a dialect encoder network. 
}
\label{fig:tts_fig}
\vspace{-5mm}
\end{figure}

Data augmentation has shown to be very effective for speech recognition (e.g.\ \cite{cui2015data,park2019specaugment}) and speaker recognition \cite{xvectors}. 
Although data augmentation for speech synthesis was investigated in the past (e.g.\ \cite{hsu2019towards,hwang2020mel}), improvements are rather small and the best augmentation strategy for speech synthesis is still unknown. 
In this paper we consider two speaker augmentation ideas and investigate how such  augmentation improves multi-speaker end-to-end TTS. 

\vspace{-1mm}
\subsection{Artificial speaker augmentation}
\vspace{-1mm}
\label{sec:artificialspeakeraugmentation}

The first method of speaker augmentation is the same as \cite{yamamoto2019speaker} wherein we create ``artificial" speakers by manipulating the high-quality audio signals. This is a re-sampling of waveforms, and the resulting signals have different fundamental frequency, speaking rate, formants, and spectra.  
We implemented this augmentation using the SoX \cite{sox} `speed' command, which speeds up or slows down audio by resampling.
We created `x0.9' and `x1.1' re-sampled versions of each VCTK speaker's speech, and used this augmented dataset to train a speaker-augmented Tacotron model.

\vspace{-1mm}
\subsection{Speaker augmentation using low-quality data}\label{sec:speakeraugmentatoinlow-quality}
\vspace{-1mm}

The second method of speaker augmentation is to use low-quality data collected for purposes other than TTS, such as ASR.  Such data can represent a diverse range of speakers and dialects, and can be used for the purpose of speaker augmentation for speech synthesis. Our aim is to use the low-quality data for speaker augmentation only, and we assume that target speaker data is limited but recorded in high-quality studios. This is different from previous work, which uses low-quality recordings for speaker adaptation \cite{hu2019neural,8683561,hsu2019hierarchical} and multi-speaker modeling (e.g.\ \cite{Zen2019}).

However, such ASR data does not meet our high-quality recording standards. It may contain background noise and reverberation unlike typical TTS data recorded in anechoeic chambers.
Furthermore, unlike artificial speaker augmentation, it also increases the number of different dialects included in the training database. We therefore modify two parts of our Tacotron TTS that use neural speaker embeddings to explicitly handle the two factors, channel and dialect, brought by low-quality data for speaker augmentation as shown in  Figure \ref{fig:tts_fig}.

\noindent \textbf{Channel-aware postnet}: The first revision is to make Tacotron's postnet dependent on channel information.  Here, the channel means all of recording equipment, noise, and reverberation. We simply use a one-hot channel label that indicates which dataset each utterance comes from during training. This channel label is input to each convolution layer of the postnet, which controls shaping and enhancement of the spectrum predicted by Tacotron's decoder.  Then, at synthesis time, we choose the highest-quality channel setting (VCTK) which will allow the model to produce speech with both a better speaker representation as well as high audio quality. This idea is relevant to \cite{8683561,hsu2019hierarchical} wherein the channel factor is used to condition the decoder. In our idea, we view Tacotron as a speech production model and re-interpret its postnet as a channel model.

\noindent \textbf{Dialect encoder network and neural dialect embeddings}:
The second revision is to make Tacotron's encoder dependent on the dialect of target speakers included in training and adaptation data.\footnote{Here dialect means English varieties. 
In traditional TTS, a lexicon and phone set for different dialects are manually prepared \cite{combilex}. 
}  
We aim to use either a common phone set or character input for all speakers, and factorise Tacotron's encoder based on neural dialect embedding vectors, computed from audio signals. Dialect identification can be considered as a subtask of spoken language recognition, and in general, approaches from speaker recognition tasks can be directly transferred to dialect identification, see~\cite{dehak2011language,garcia2016stacked,cai2018novel,snyder2018spoken}. Therefore, similar to our speaker encoder network, we reused the Learnable Dictionary Encoding (LDE) \cite{cai2018exploring} based network architecture for our dialect encoder network. For more details, refer to Section 2 of~\cite{cooper2020zeroshot}.

\vspace{-1mm}
\section{Experiments}
\vspace{-1mm}
    \subsection{Setup}
    \vspace{-1mm}

    We use two baseline models, a phoneme-based model which is the same as the best system from our prior work \cite{cooper2020zeroshot}, and one with character input.  
    4 speakers are held out as validation data and 4 speakers are held out as the test set.   80-dimensional mel spectrograms that are output from Tacotron are converted to 16 kHz waveforms using WaveNets \cite{wavenet} that were trained on the same VCTK training set. Details of the setup can be found in Section 3 and 4 of~\cite{cooper2020zeroshot}, and code is available online\footnote{\href{https://github.com/nii-yamagishilab/multi-speaker-tacotron}{https://github.com/nii-yamagishilab/multi-speaker-tacotron}}.
    
\begin{table}[tb]
\centering
      \caption{Number of speakers in training, validation and test sets}
      \label{tab:speaker}
      \scriptsize
      \vspace{-1mm}
      \begin{tabular}{@{}llccc@{}}
        \toprule
        Models & Data type & Train & Dev & Test \\
        \midrule
        Baseline & VCTK & 100 & 4 & 4 \\ 
        \midrule
        Baseline    & VCTK & 100 & 4 & 4 \\ 
        +artificial & VTLP & 200 & 8 & - \\ 
        \midrule
        Baseline     & VCTK    & 100 & 4 & 4 \\ 
        +low-quality & non-TTS & 200 & 8 & - \\ 
        \bottomrule
      \end{tabular}
      \vspace{-5mm}
\end{table}
    
    \vspace{-1mm}
    \subsection{Artificial speaker augmentation}
    \vspace{-1mm}

    We created an augmented VCTK dataset by speeding up and slowing down the speech of each original VCTK speaker as described in Section \ref{sec:artificialspeakeraugmentation} and giving them unique speaker identities, resulting in three times as many ``speakers'' as in the original dataset.  Then, we trained both character-based and phone-based models in the same manner as our baselines except using the larger augmented dataset. 

    \vspace{-1mm}
    \subsection{Speaker augmentation using low-quality ASR corpora}\label{sec:speakeraugmentatoinlow-quality-exp}
    \vspace{-1mm}

    We create a large mixed dataset for TTS  training using both VCTK and a variety of corpora collected for ASR, which contain a variety of recording conditions and English dialects.  While we hold out some portion of each corpus for validation and test, we focus our actual evaluation on VCTK speakers.  Once again we train both character-input and phone-input models.  We used standard train/validation/test sets where they were defined, as well as predefined adaptation utterances or utterances that were common across speakers for extracting speaker embeddings.  We kept the number of training speakers the same as in the artificially-augmented VCTK set.  Two speakers were chosen per corpus to add to our development set for the purposes of preliminary model evaluation and selection. Below, we briefly describe the four ASR corpora used in our multi-speaker TTS training (with info about number of speakers in Table \ref{tab:speaker}):
    
    \vspace{1mm}
    \noindent \textbf{GRID} \cite{grid}:  This corpus consists of 32 English speakers (15 training set speakers) speaking English, Scottish, and Jamaican dialects.  Sentences are all of the form ``place green at B 4 now.''  While all sentences are technically unique, they are each very similar (many only varying by one word) and the vocabulary is small. There are 1000 utterances per speaker.  Audio is 16 bit and 50 kHz.  Some recordings contain small amounts of background noise such as mouse clicks.
    
    \vspace{1mm}
    \noindent \textbf{WSJ1} \cite{wsj1}: Wall Street Journal read by speakers of various American English dialects.  Audio is 16 bit at a 16kHz sampling rate.  We used the first 50 of the 200 `si\_tr\_s' training set speakers, who each have around 200 utterances.
    
    \vspace{1mm}
    \noindent \textbf{WSJCAM} \cite{wsjcam}:  Wall Street Journal sentences read by speakers of various British English dialects.  Audio  is 16 bit at a 16kHz sampling rate.  We used 85 of the 96 training speakers, who each read about 110 sentences.  Recordings contain loud audible line noise and reverberation.
    
    \vspace{1mm}
    \noindent \textbf{TIMIT} \cite{timit}:  Speakers of eight American English dialects each read ten phonetically-rich sentences.  We picked 50 of the 462 training speakers, balancing for gender and dialect.  Audio is 16 bit at a 16kHz sampling rate.
    
    \vspace{-1mm}
    \subsection{Modeling channel and dialect factors}
    \vspace{-1mm}
    \noindent \textbf{Ground-truth channel labels}: In addition to training directly by mixing VCTK with the four new ASR corpora, we also trained phone and character models provided with ground-truth channel labels.  
    We used a one-hot encoding indicating which corpus each training utterance comes from\footnote{In addition to the one-hot code, we also tried a binary code simply representing TTS data (VCTK) or not (all other corpora), but this resulted in worse development set alignment error rates.}, and channel labels are input to the Tacotron postnet. 
    
    \noindent \textbf{LDE-based neural dialect embeddings}: 
    Given our goal of modeling English dialects only, using the standard NIST LRE recipe is not ideal\footnote{\href{https://github.com/kaldi-asr/kaldi/tree/master/egs/lre07}{https://github.com/kaldi-asr/kaldi/tree/master/egs/lre07}}.
    We opted to use the ATR dialect corpus\footnote{\href{https://www.atr-p.com/products/sdb.html}{https://www.atr-p.com/products/sdb.html}} with six English dialects: Australian, British and various American English.
    Read and spontaneous speech recordings are sampled such that they are balanced for training.
    Our dialect encoder network is based on LDE, and we performed a hyper-parameter sweep.
    Similar to the speaker embeddings in ~\cite{cooper2020zeroshot}, we computed the cosine-similarity scores between dialect embeddings of the synthesized and ground-truth speech\footnote{We want to emphasize that our strategy is not optimal, and a strong assumption we imposed here is that the cosine-similarity and speaker/dialect distributions is a one-to-one mapping.}, and accordingly selected five best embeddings \textit{each} for phone and character models.  Details of these embeddings are in Table \ref{tab:DE}.

    \begin{table}[tb]
\centering
      \caption{5 best dialect embeddings (DE) for phone- and character-based TTS.  Number of dimensions, mean-only (m) or mean and standard deviation (m,s) pooling, and the number of dictionary components in the pooling layer are shown.}
      \label{tab:DE}
      \scriptsize
        \vspace{-1mm}
      \begin{tabular}{@{}l|lll|lll@{}}
        \toprule
          & \multicolumn{3}{c|}{Phone} & \multicolumn{3}{c}{Char} \\
          & dim & pl & dc & dim & pl & dc \\
        \hline
        DE1    & 256 & m,s & 32 & 128 & m,s & 32 \\ 
        \hline
        DE2    & 256 & m   & 64 & 256 & m   & 32 \\ 
        \hline
        DE3    & 256 & m,s & 64 & 32  & m,s & 64 \\ 
        \hline
        DE4    & 32  & m,s & 64 & 512 & m,s & 32 \\ 
        \hline
        DE5    & 64  & m   & 64 & 64  & m,s & 32 \\ 
        \bottomrule
      \end{tabular}
      \vspace{-5mm}
\end{table}

\begin{figure}[tb] 
    \centering
    \includegraphics[width=1.0\linewidth]{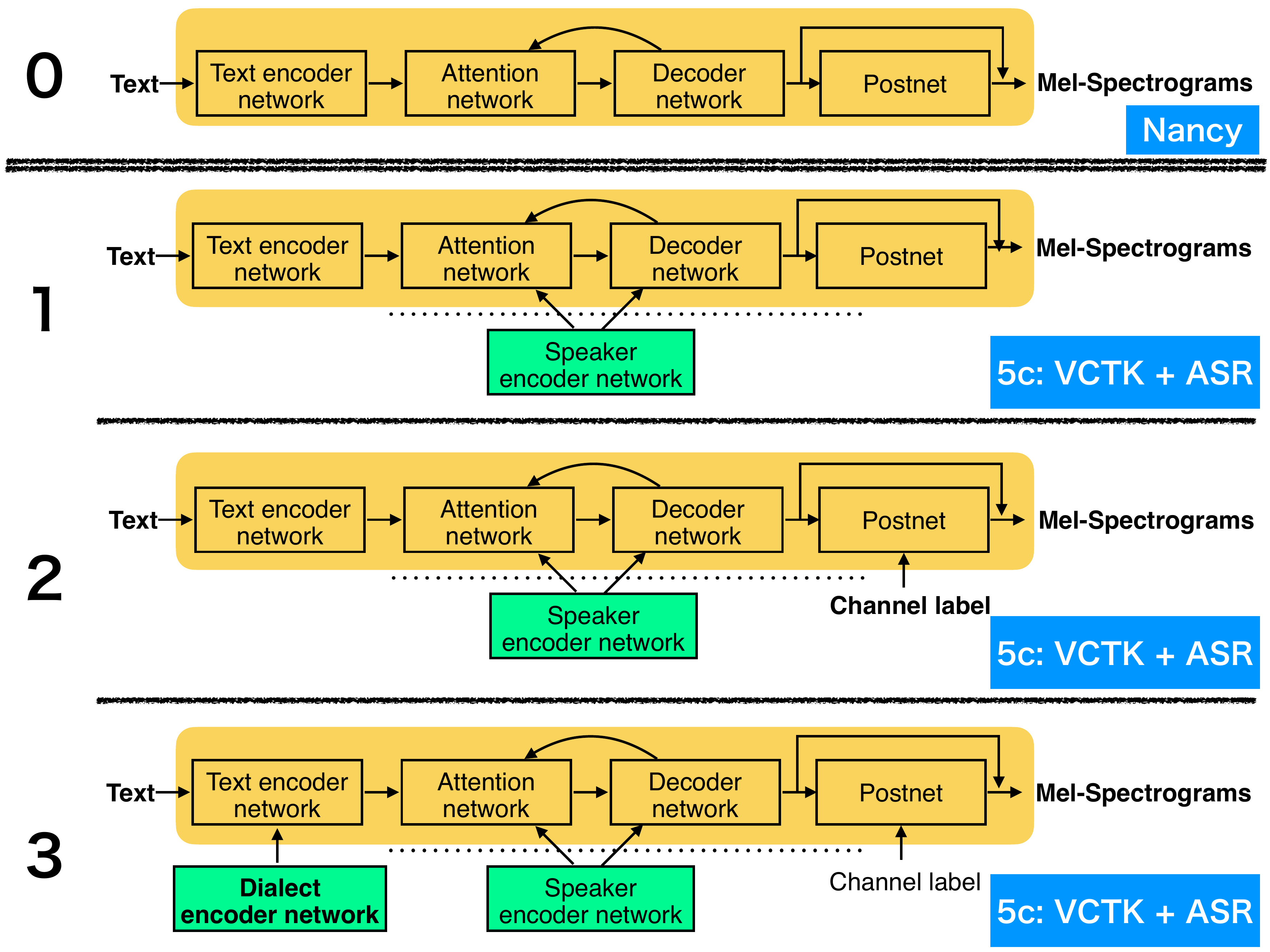}
    \vspace{-3mm}
    \caption{Illustration of the warmstart training strategy for our Tacotrons (w/o a WaveNet) in this work. \textbf{Green} denotes the pretrained components, \textbf{Yellow} denotes end-to-end training, and \textbf{Blue} denotes the training data.}
    \label{fig:warmstart_fig}
    \vspace{-5mm}
    \end{figure}

    \noindent \textbf{Warm-start training strategy\footnote{We found this strategy effective, as it produces better synthesis quality and reduces training time.}}: 
    We adopted a warm-start training scheme, in which the full Tacotron training is broken down to four phases (see Figure~\ref{fig:warmstart_fig}) where the parameters in each phase are initialized from that of previous phase.
    In Phase 0, a seed single-speaker Tacotron2 is trained on the Nancy dataset from Blizzard 2011~\cite{blizzard2011}.
    In Phase 1, we trained a multi-speaker gender-dependent model on 5 corpora (VCTK + ASR), with parameters initialized from the previous step, and included speaker embeddings extracted from a separately-trained LDE model with mean pooling and angular softmax, trained on VoxCeleb~\cite{nagrani2017voxceleb,chung2018voxceleb2}. These embeddings are concatenated with the encoder output and input to the attention mechanisms, as well as input to the prenet to the decoder~\cite{cooper2020zeroshot}.
    In Phase 2, we added channel labels.  In Phase 3, finally, using one of the top 5 dialect embeddings for phone or character models, we continued training with all five corpora, channel labels, and speaker and dialect embeddings.
    Each phase is trained until convergence.
    
\vspace{-1mm}
\subsection{Subjective evaluation setup}
\vspace{-1mm}

 We conducted a crowdsourced online listening test with native English listeners.  We asked listeners to rate each sample on a mean opinion score (MOS) scale of 1-5 for naturalness and on a differential MOS (DMOS) scale of 1-5 for speaker similarity compared to a ground truth sample from the target speaker.  We also asked listeners to provide a categorical opinion about dialect from six choices: American, Canadian, English, Irish, Northern Irish, and Scottish.  Since listeners may be unfamiliar with these accents, we also provided reference samples of each accent from VCTK speakers who were not included in the test, on a separate webpage that listeners may optionally refer to.  We evaluated 20 different systems: natural speech, vocoded speech using WaveNet, phone and character baselines, VTLP-augmented models, models trained with additional ASR data for a total of 5 training corpora (5c), models with 5c and channel label (CL), and models with 5c + CL + dialect embeddings (DE). For each system, we generated 20 samples using text that was unseen during training from each of 4 VCTK training set (seen) speakers, 4 development set speakers, and 4 test set speakers (completely unseen).  We grouped samples into sets of of 40 utterances each, and had 5 different listeners evaluate each set.  A total of 60 listeners completed the test, rating 10 sets each.  

\noindent \textbf{Metric for evaluating dialect confusion}: Since dialect ID can be a challenging task even for native listeners, we evaluated confusion matrices of true vs.\ guessed accents.  We computed Frobenius distance \cite{amendola2015model,laurent2012forecasting} between the confusion matrix for dialects of natural speech and those for each TTS system, based on the idea that if a confusion matrix for TTS is similar to the one for natural speech, then accents are well-represented.

\vspace{-1mm}
\subsection{Subjective evaluation results and analysis}
\vspace{-1mm}

MOS and DMOS results are presented in Table \ref{tab:subj}.  Statistical significances were measured using the Mann-Whitney U test at a threshold of p=0.01, and systems are compared with their {\em respective} baselines, i.e. phone or character.  Significantly better and worse systems are highlighted in red and blue, respectively.

        \begin{table}[t!]
          \caption{\it Results: MOS and DMOS on a scale of 1-5 for seen (train) and unseen (dev and test) speakers. Synthesis was done using unseen texts. Here, 5c denotes the 5 training corpora (VCTK + 4 ASR), CL denotes channel label, and DE\{1..5\} denotes the 5 best dialect embeddings for char and phone models. Significant improvements over the baseline are highlighted in red, and significantly worse systems are in blue.} 
          \label{tab:subj}
          \centering
          \scriptsize 
            \begin{tabular}{@{}|l||c|c|c||c|c|c|@{}}
            \hline
             & \multicolumn{3}{c||}{Naturalness} & \multicolumn{3}{c|}{Speaker Similarity} \\
            \hline
            system & train & dev & test & train & dev & test \\
            \hline
            \hline
            natural & 4.5 & 4.4 & 4.5 & 4.6 & 4.5 & 4.5 \\
            \hline
            \hline
            vocoded & 4.2 & 4.2 & 4.3 & 4.1 & 3.9 & 4.0 \\
            \hline
            \hline
            \cellcolor{green!25}phone baseline & 3.6 & 3.7 & 4.0 & 3.7 & 2.0 & 2.7 \\
            \hline
            \cellcolor{green!25}phone VTLP & 3.7 & 3.7 & 4.0 & 3.6 & 2.1 & 2.7 \\
            
            \hline
            \cellcolor{green!25}phone 5c & 3.8 & \cellcolor{blue!25} 3.5 & \cellcolor{blue!25} 3.7 & \cellcolor{blue!25} 3.4 & 2.1 & 2.6 \\
            \hline
            \cellcolor{green!25}phone 5c+CL         & 3.7                     & 3.7                     & 3.9                     & 3.5                     & 2.0                    & 2.5 \\
            
            \hline
            \cellcolor{green!25}phone 5c+CL+DE1 & \cellcolor{blue!25} 2.4 & \cellcolor{blue!25} 2.4 & \cellcolor{blue!25} 2.5 & \cellcolor{blue!25} 3.2 & \cellcolor{red!25} 2.3 & \cellcolor{blue!25} 2.4 \\
            \hline
            \cellcolor{green!25}phone 5c+CL+DE2 & \cellcolor{blue!25} 2.5 & \cellcolor{blue!25} 2.4 & \cellcolor{blue!25} 2.5 & \cellcolor{blue!25} 3.3 & \cellcolor{red!25} 2.2 & 2.6 \\
            
            \hline
            \cellcolor{green!25}phone 5c+CL+DE3 & \cellcolor{red!25} 3.9 & 3.7 & 3.9 & 3.6 & 2.1 & 2.6 \\
            \hline
            \cellcolor{green!25}phone 5c+CL+DE4 & \cellcolor{blue!25} 1.1 & \cellcolor{blue!25} 1.1 & \cellcolor{blue!25} 1.1 & \cellcolor{blue!25} 2.8 & \cellcolor{red!25} 2.4 & 2.5 \\
            
            \hline
            \cellcolor{green!25}phone 5c+CL+DE5 & \cellcolor{red!25} 3.9 & 3.7 & \cellcolor{blue!25} 3.8 & \cellcolor{blue!25} 3.4 & 2.0 & 2.6 \\            
            \hline
            \hline
            
            \cellcolor{yellow!40}char baseline & 3.7 & 3.7 & 3.9 & 3.6 & 1.9 & 2.8 \\
            \hline
            \cellcolor{yellow!40}char VTLP & 3.8 & 3.6 & 3.9 & 3.5 & 2.1 &  2.5 \\
            
            \hline
            \cellcolor{yellow!40}char 5c & 3.7 & \cellcolor{blue!25} 3.5 &  \cellcolor{blue!25} 3.7 & \cellcolor{blue!25} 3.3 & 2.0 & 2.6 \\
            \hline
            \cellcolor{yellow!40}char 5c+CL & 3.8 & 3.8 & 3.9 & 3.6 & 2.1 & 2.6 \\
            
            \hline
            \cellcolor{yellow!40}char 5c+CL+DE1 & \cellcolor{blue!25} 2.5 & \cellcolor{blue!25} 2.5 & \cellcolor{blue!25} 2.5 & \cellcolor{blue!25} 3.3 & 2.1 & \cellcolor{blue!25} 2.5 \\
            \hline
            \cellcolor{yellow!40}char 5c+CL+DE2 & \cellcolor{red!25} 4.0 & 3.7 & 4.0 & 3.6 & 2.0 & 2.5 \\
            
            \hline
            \cellcolor{yellow!40}char 5c+CL+DE3 & \cellcolor{red!25} 3.9 & \cellcolor{blue!25} 3.4 & 3.8 & 3.6 & 2.1 & \cellcolor{blue!25} 2.4 \\
            \hline
            \cellcolor{yellow!40}char 5c+CL+DE4 & \cellcolor{red!25} 4.0 & 3.7 & 3.9 & 3.6 & 2.1 & 2.5 \\
            
            \hline
            \cellcolor{yellow!40}char 5c+CL+DE5 & \cellcolor{red!25} 3.9 & 3.8 & 3.9 & 3.5 & 2.0 & 2.6 \\            
            \hline            
          \end{tabular}
          \vspace{-5mm}
        \end{table}
        
        \noindent 
        \textbf{A) Baseline vs. speaker augmentation:} The MOS and DMOS results show an unexpected but interesting tendency. Contrary to our initial expectation, we obtained statistically significant improvements for naturalness of {\em seen} speakers when the low-quality data, both channel-aware postnet, and dialect-aware encoder are all used for Tacotron training. This is surprising in two ways. First, speaker augmentation contributes to  naturalness rather than speaker similarity. Second, adding low-quality data paradoxically resulted in improved quality of synthetic speech for seen speakers. This is somewhat surprising, but this phenomenon has been clearly confirmed for 2 phone-based systems and 4 character-based systems. 
        MOS scores for the phone and character systems have been increased from 3.6 to 3.9 and from 3.7 to 4.0, respectively.  Speaker similarity of the development set speakers was also improved from 2.0 to 2.4 for some of the phone-based systems.  We may speculate that the addition of dialect modeling and a larger variety of different speakers helps to  capture important aspects of speech, but that overfitting to speakers seen during training is still taking place.
        
        \noindent 
        \textbf{B) Artificial vs.\ low-quality data:} Next, we see that VTLP (artificial speaker augmentation) did not improve naturalness or speaker similarity, although it is known that this method works well in other tasks. On the other hand, mixing non-ideal data carelessly does worsen results: we see that simply mixing low-quality data produces significantly worse results in some cases, and that adding the channel-aware postnet only shows improvement when combined with dialect embeddings.  This indicates that we need to handle both channel and dialect factors properly.
         
        \noindent 
        \textbf{C) Impacts of dialect encoders:} One implication from Table~\ref{tab:subj} is that the effect of different types of dialect encoders on synthesis is unclear and including them does not consistently improve naturalness and speaker similarity.  However, they do appear to be necessary for better dialect modeling (see Table~\ref{tab:frob}).

            \begin{table}[t!]
          \caption{\it Frobenius distance results for dialects of seen (train) and unseen (dev and test) speakers, compared to confusion matrices for dialects of natural speech.  Distances smaller than baseline are highlighted in red, with best result per category in bold.  Distances larger than baseline are highlighted in blue.} \label{tab:frob}
          \centering
          \scriptsize
            \begin{tabular}{@{}|l||c|c|c|@{}}
            \hline
             & 
             \multicolumn{3}{c|}{Dialect confusion} \\
            \hline
            system & train & dev & test \\
            \hline
            \hline
            vocoded & 0.06 & 0.32 & 0.32 \\
            \hline
            \hline
            \cellcolor{green!25}phone baseline & 0.20 & 1.06 & 1.12 \\
            \hline
            \cellcolor{green!25}phone VTLP & \cellcolor{blue!25} 0.31 & \cellcolor{red!25} 0.86 & \cellcolor{blue!25} 1.20 \\
            
            \hline
            \cellcolor{green!25}phone 5c & \cellcolor{red!25} 0.19 & \cellcolor{red!25} 0.93 & \cellcolor{red!25} 0.93 \\
            \hline
            \cellcolor{green!25}phone 5c+CL & \cellcolor{red!25} 0.19 & \cellcolor{red!25} \bf 0.84 & \cellcolor{red!25} 0.99 \\
            
            \hline
            \cellcolor{green!25}phone 5c+CL+DE1 & \cellcolor{blue!25} 0.42 & \cellcolor{red!25} 0.84 & \cellcolor{red!25} 0.88 \\
            \hline
            \cellcolor{green!25}phone 5c+CL+DE2 & \cellcolor{blue!25} 0.34 & \cellcolor{red!25} 0.95 & \cellcolor{red!25} 0.81 \\
            
            \hline
            \cellcolor{green!25}phone 5c+CL+DE3 & \cellcolor{red!25} \bf 0.13 & \cellcolor{red!25} 0.93 & \cellcolor{red!25} 0.95 \\
            \hline
            \cellcolor{green!25}phone 5c+CL+DE4 & \cellcolor{blue!25} 0.44 & \cellcolor{red!25} 0.88 & \cellcolor{red!25} 0.90 \\
            
            \hline
            \cellcolor{green!25}phone 5c+CL+DE5 & \cellcolor{blue!25} 0.20 & \cellcolor{red!25} 0.92 & \cellcolor{red!25} \bf 0.79 \\            
            \hline
            \hline
            
            \cellcolor{yellow!40}char baseline & 0.25 & 0.96 & 0.86 \\
            \hline
            \cellcolor{yellow!40}char VTLP & \cellcolor{red!25} \bf 0.12 & \cellcolor{blue!25} 1.00 & \cellcolor{blue!25} 1.17 \\
            
            \hline
            \cellcolor{yellow!40}char 5c & 0.25 & 0.96 & 0.86 \\
            \hline
            \cellcolor{yellow!40}char 5c+CL & 0.25 & \cellcolor{red!25} \bf 0.86 & \cellcolor{red!25} \bf 0.79 \\
            
            \hline
            \cellcolor{yellow!40}char 5c+CL+DE1 & \cellcolor{blue!25} 0.41 & \cellcolor{red!25} 0.91 & \cellcolor{red!25} 0.83 \\
            \hline
            \cellcolor{yellow!40}char 5c+CL+DE2 & \cellcolor{red!25} 0.17 & \cellcolor{red!25} 0.92 & \cellcolor{blue!25} 1.33 \\
            
            \hline
            \cellcolor{yellow!40}char 5c+CL+DE3 & \cellcolor{red!25} 0.18 & \cellcolor{red!25} 0.92 & \cellcolor{blue!25} 1.02 \\
            \hline
            \cellcolor{yellow!40}char 5c+CL+DE4 & \cellcolor{red!25} 0.21 & \cellcolor{red!25} 0.91 & \cellcolor{blue!25} 1.15 \\
            
            \hline
            \cellcolor{yellow!40}char 5c+CL+DE5 & \cellcolor{red!25} 0.22 & \cellcolor{blue!25} 1.02 & \cellcolor{blue!25} 1.08 \\            
            \hline            
          \end{tabular}
          \vspace{-5mm}
        \end{table}

        \noindent 
        \textbf{D) Dialect identification and confusion:}
        Frobenius distances representing confusions of perceived dialects are shown in Table \ref{tab:frob}. The Frobenius distance means how similar confusion matrices of perceived dialects of synthetic speech are compared to those of natural speech. We observed relative improvements for unseen speakers (dev and test). All phone-based systems using the low-quality data have smaller Frobenius distances than the baseline system for unseen speakers. This means adding low-quality data helps our synthesizers generate appropriate phones better and to better match dialects correctly with respect to listeners' perception.  It also helps some of the character-based systems to use channel-aware postnet and the dialect-aware encoder. On the other hand, we see that unseen speakers (dev and test) have much larger Frobenius distances compared to seen speakers, even for vocoded speech. This tendency is consistent with speaker similarity judgements in Table \ref{tab:subj}. 

\vspace{-1mm}
\section{Conclusions}
\vspace{-1mm}

In this paper we investigated two realistic speaker augmentation scenarios for multi-speaker end-to-end speech synthesis: artificial augmentation and the use of non-ideal low-quality data. We revised the postnet and encoder of Tacotron to support channel and dialect variations from the low-quality data.
Experimental results revealed that using low-quality data with various English accents is an effective data augmentation method for multi-speaker end-to-end speech synthesis.
Contrary to our initial expectations, naturalness of seen speakers has been improved and listeners' ratings of perceived dialects are better matched to natural speech for unseen speakers.
Our results suggest that improving speaker similarity still remains a challenge, and future work includes the use of large low-quality databases for training an initial seed model and fine-tuning it to a high-quality corpus. 

\vspace{1mm}

\noindent
\textbf{Acknowledgments}
\footnotesize{CI is supported by the Merrill Lynch Fellowship, MIT. This work was partially supported by a JST CREST Grant (JPMJCR18A6, VoicePersonae project), Japan, and by MEXT KAKENHI Grants (16H06302, 18H04112, 18KT0051, 19K24372), Japan. The numerical calculations were carried out on the TSUBAME 3.0 supercomputer at the Tokyo Institute of Technology.  We thank Yi Zhao for help with Frobenius distance, Jim Glass and Hao Tang for their discussions.}



\bibliographystyle{IEEEtran}
\bibliography{main}
\end{document}